\documentclass[aps,prb,reprint,groupedaddress,superscriptaddress]{revtex4-2}
\usepackage[T1]{fontenc}
\usepackage{amssymb, amsmath, bbold, mathbbol}
\usepackage{graphicx}
\usepackage[usenames,dvipsnames]{color}
\usepackage[table]{xcolor}
\usepackage{physics}
\usepackage[normalem]{ulem}
\usepackage{soul}
\usepackage{hyperref}

\usepackage{subcaption}

\newcommand{\be}{\begin{equation}}
\newcommand{\ee}{\end{equation}}
\newcommand{\bea}{\begin{eqnarray}}
\newcommand{\eea}{\end{eqnarray}}

\newcommand{\m}{\mathrm}

\begin{document}

\title{The crucial role of substrate in FeSe/STO: new insights to interface-driven superconductivity from first-principles
}
\author{R. Reho}
\affiliation{Chemistry Department and Debye Institute for Nanomaterials Science, Condensed Matter and Interfaces, Utrecht University and ETSF, PO Box 80.000, 3508 TA Utrecht, The Netherlands}
\email{r.reho@uu.nl}

\author{N. Wittemeier}
\affiliation{Catalan Institute of Nanoscience and Nanotechnology (ICN2) and European Theoretical Spectroscopy Facility, CSIC, BIST, Campus UAB, Bellaterra, 08193 Barcelona, Spain}
\email{nils.wittemeier@icn2.cat \newline RR and NW contributed equally to this work.}

\author{A. H. Kole}
\affiliation{Chemistry Department and Debye Institute for Nanomaterials Science, Condensed Matter and Interfaces, Utrecht University and ETSF, PO Box 80.000, 3508 TA Utrecht, The Netherlands}
\email{a.h.kole@uu.nl}

\author{A. R. Botello-M\'endez}
\affiliation{Chemistry Department and Debye Institute for Nanomaterials Science, Condensed Matter and Interfaces, Utrecht University and ETSF, PO Box 80.000, 3508 TA Utrecht, The Netherlands}
\email{a.r.botellomendez@uu.nl}

\author{Zeila Zanolli}
\affiliation{Chemistry Department and Debye Institute for Nanomaterials Science, Condensed Matter and Interfaces, Utrecht University and ETSF, PO Box 80.000, 3508 TA Utrecht, The Netherlands}
\email{z.zanolli@uu.nl}
\date{\today}

\begin{abstract}
We investigate the superconducting properties of monolayer FeSe, both freestanding (ML FeSe) and on SrTiO$_3$ (STO), by simultaneously solving the Kohn-Sham Density Functional Theory and Bogoliubov--de Gennes equations. 
Our results demonstrate that the substrate profoundly alters both the normal-state and superconducting properties of FeSe. 
We identify proximity-induced superconductivity in the interfacial TiO$_2$ layer of STO, due to hybridization between Fe $d$ and O $p$ orbitals. 
This hybridization results in a fivefold increase in the superconducting gap width and confines superconducting states to the $M$ point in the Brillouin Zone. This is in contrast to ML FeSe, where superconductivity emerges at both the $\Gamma$ and $M$ points. 
Furthermore, the substrate modifies the orbital character of the states responsible for superconductivity, which change 
from Fe $d_{z^2}$ in ML FeSe to Fe $d_{xz}/d_{yz}$ in FeSe/STO.  
In both systems, we demonstrate an anisotropic superconducting gap with multiple coherence peaks, originating at different k-points in the Brillouin Zone.
Additionally, in FeSe/STO, we identify emerging states unique to the superconducting phase arising from electron-hole hybridization at $M$, in agreement with experiments.
Our findings highlight the decisive impact of substrate (hybridization, strain, charge transfer, magnetic order)  on the superconducting properties of FeSe. We suggest potential pathways for engineering novel high-temperature FeSe-based superconductors by leveraging interfacial interactions in substrates with high electron affinity.
\end{abstract}
\maketitle

\section*{Introduction}
Monolayer (ML) FeSe on SrTiO$_3$ (FeSe/STO) features high critical temperature (T$_C$$\sim$ 40--100~K) making it appealing for both fundamental research and development of next-generation quantum devices~\cite{wang2012interface,huang2017monolayer,zhang2014interface}.
The electronic dispersion of FeSe is highly sensitive to factors such as magnetic configurations, structural changes (strain) and the dielectric environment (doping and dipole interactions)\cite{zheng2013antiferromagnetic}.
The presence of a substrate affects all the above properties which, in turn,
directly influence the superconducting pairing mechanism~\cite{coldea2018key}.
Experiments on FeSe/STO proved the conventional BCS--like $s$--wave nature of the superconducting pairing potential~\cite{wei2024particle}. 
{\it Zhang et al.}~\cite{zhang2016superconducting} reported two anisotropic superconducting gaps driven by Fe $d_{xz}/d_{yz}$  ($20$~meV wide) and Fe $d_{xy}$ ($26$~meV wide),
reflecting the multiorbital nature of the Fermi surface and the orbital--dependent pairing in iron chalcogenides.
Currently, it is still unclear whether FeSe/STO exhibits single or double coherence peaks~\cite{liu2015electronic}, the interplay between its magnetic and electronic properties is not fully understood, the agreement between theoretical predictions and experiments is incomplete, and the origin of high T$_C$ is unknown.

Tight-binding models of FeSe ML and FeSe/STO successfully describe specific experiments but lack general applicability~\cite{liu2015electronic, wang2012interface} and are restricted to a normal state description.
State-of-the-art theoretical studies ~\cite{acharya2021electronic, acharya2022role, acharya2023vertex} combine dynamical mean field theory (DMFT) and the quasiparticle self-consistent GW approximation to describe the superconducting instability of FeSe/STO from the spin and charge response of the system. 
These studies explain the dependence of the critical temperature on doping and Fe-O interaction.
This method includes electron correlations and requires careful parameter tuning. 

In this work, we simultaneously solve the Kohn-Sham Density Functional Theory (DFT) and Bogoliubov–de Gennes (BdG) equations~\cite{reho2024density, garcia2020siesta}
to analyze and interpret both the normal and superconducting states of ML FeSe and FeSe/STO. 
Our findings highlight the significant impact of the substrate on the superconducting (SC) phase of FeSe, and emphasize how interface-driven superconductivity can be leveraged to enhance superconductivity in FeSe. 
In the normal state, metallicity in ML FeSe arises from an electron pocket a $M$ and a hole pocket along the $\Gamma-X$. In contrast, in FeSe/STO, the hole pocket shifts to lower energies within the valence band.
This difference influences the mechanism behind superconductivity.
In the superconducting phase, we predict proximity-induced superconductivity on the interfacial $O$ atoms and a fivefold widening of the SC gap in FeSe/STO compared to ML FeSe. 
We elucidate the origins of superconducting coherence peaks in the superconducting density of states (SC--DOS) of both system, identifying univocally the real-space orbitals and momentum vector $\mathbf{k}$ associated with these peaks.
We characterize the Bogoliubov–de Gennes spectrum and predict an anisotropic, momentum-dependent superconducting gap $\Delta(\mathbf{k})$ for both system.
We analyze the changes in electronic dispersion between freestanding ML FeSe and FeSe/STO, identifying the dominant roles of Fe $d$ and O $p$ orbitals in shaping the Fermi surface.  
We conclude that FeSe and FeSe/STO act as fundamentally different materials, as substrate hybridization, interfacial interactions and charge transfer play a crucial role in shaping the microscopic superconducting mechanism. This behaviour can be generalized to the whole family of iron pnictides. Substrate engineering presents a promising approach to design superconductors with enhanced $T_C$.

\section*{Results}

\subsection*{Structural properties}
In the following, we focus on the tetragonal phase of ML FeSe, in view of comparing it with Fe/STO~\footnote{While single crystal FeSe undergoes a tetragonal-to-orthorhombic phase transition below 90K~\cite{coldea2018key}, we expect that the symmetry of the STO surface (D$_4$) will stabilize the tetragonal FeSe phase of FeSe in FeSe/STO~\cite{coldea2018key}}.
The primitive unit cell of FeSe consists of two Fe and two Se atoms. The Fe atoms form two tetragonal sub--lattices, offset relative to each other [Fig.~\ref{fig:fese_ml_sc_state}.(a)].
The Se atoms form tetrahedra centered on the Fe atoms.
After structural relaxation, the lattice parameter of ML FeSe is $3.71$ \AA~ and the Se-Se distance is $1.44$ \AA. 
FeSe is usually grown by molecular beam epitaxy (MBE) on the TiO$_2$ terminated surface of STO~\cite{liu2015electronic}.
For this reason, we first modeled STO as a six--layer slab and performed a full relaxation (atoms and unit cell), obtaining a lattice constant of 3.92~\AA. 
The atomic positions of the FeSe/STO heterostructure were then relaxed while keeping the STO lattice fixed, resulting in 5.7\% strain on FeSe ML.
The FeSe/STO heterostructure was constructed by placing 
the bottom Se atom of FeSe directly above the top Ti atom of the STO substrate. After relaxation, the Se-Ti distance 
was $2.95$~\AA ~[Fig.~\ref{fig:fese_sto_mainplot}.(a)].

\subsection*{Magnetic instabilities} 
\begin{figure}[h]
    \centering
    \includegraphics[width=1.\columnwidth]{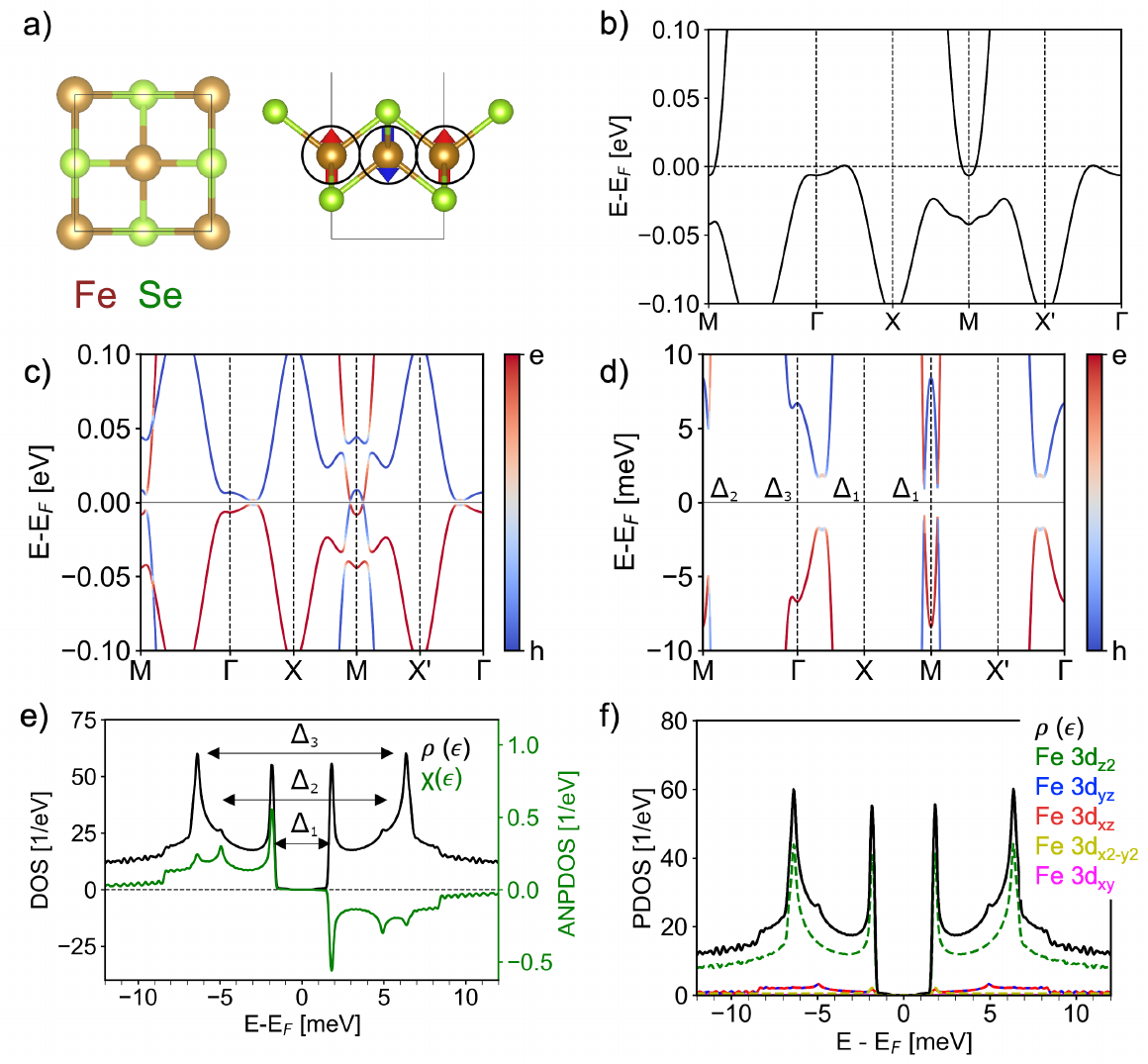}
    \caption{ML FeSe in the CB-AFM magnetic phase. (a) Top and side view of a unit cell of ML FeSe, with red and blue arrows representing spin directions and black circles illustrating the initialization of the superconducting pairing potential. (b) Normal state band structure. (c) Superconducting band structure projected on the electron (red) and hole (blue) components of the BdG spectrum. In (d) we report a narrower energy range around the SC gap. 
    (e) SC--DOS ($\rho(\varepsilon)$, black) and anomalous SC--DOS ($\chi(\varepsilon)$, green)  exhibiting superconducting gaps with widths $\Delta_1 = 3.6$, $\Delta_2 = 10.0$, and $\Delta_3 = 12.8$ meV. (f) SC--DOS projected over Fe orbitals.}
    \label{fig:fese_ml_sc_state}
\end{figure}

FeSe, both in its bulk and ML forms, exhibits a complex energy landscape characterized by multiple competing magnetic configurations.
The lowest-energy magnetic state of bulk FeSe is theoretically predicted to be the staggered dimer phase~\cite{GlasbrennerEffectMagnetic2015}. However, long--range magnetic order is not observed in experiments. Indeed, bulk FeSe is a nematic quantum--disordered paramagnet, interpolating between the checkerboard  antiferromagnetic (CB-AFM) and stripe AFM phases, both in the low (T = 4~K) and high temperature (T = 110~K) regimes~\cite{wang2016magnetic}.
For bulk and ML FeSe, we computed first-principles total energies of various magnetic configurations (Appendix~\ref{app:magnorder}). 
In both cases, we found three low-energy competing magnetic configurations: staggered trimer, staggered dimer, and single stripe, followed by two CB-AFM phases
(in- and out-of-plane) with similar total energies, approximately $68$~meV above the lowest-energy configurations.
We find that only the band structure of the out-of-plane CB phase
[Fig.~\ref{fig:fese_ml_sc_state}.(a)] resembles the  
ARPES electronic dispersion~\cite{wang2016topological,zheng2013antiferromagnetic}. For this reason, we focus on the CB phase despite its higher total energy compared to the other phases. 

\subsection*{Normal State Electronic Properties}
The normal state band structure of ML FeSe features an electron pocket at $M$,
a hole pocket along the $\Gamma-X$ high-symmetry line, 
and a relative maximum in the highest occupied valence band along the  $\Gamma-M$ line [Fig.~\ref{fig:fese_ml_sc_state}.(b)]. 
Analysis of the Mulliken populations~\cite{mulliken1955electronic} reveals that FeSe acquires 0.13 electrons (per unit cell) from STO in the FeSe/STO heterostructure. 
As a result, the hole pocket along the $\Gamma-X$ direction is pushed below the Fermi-level, and the top valence band becomes flatter [Fig.~\ref{fig:fese_sto_mainplot}.(b)]. 

The origin of metallicity in FeSe changes due to the interaction with the substrate: 
in the freestanding ML, metallicity is associated with the electron and a hole pocket, while in FeSe/STO only to the electron pocket at $M$. 
Analysis of the Fermi surfaces (Fig.~\ref{fig:fermi_surfaces}) facilitate the visualization of these concepts:
In FeSe ML, the hole pocket along $\Gamma-X$ is clearly visible in Fig.~\ref{fig:fermi_surfaces}.(a), while the relative maxima along $\Gamma-M$ lies tens of meV lower in energy. In FeSe/STO, instead, the hole pocket near $\Gamma$ disappears. At $M$, the electron pocket of FeSe/STO has a more pronounced hyperbolic shape with respect to FeSe ML [Fig.~\ref{fig:fermi_surfaces}.(b)].
This will have consequences in the superconducting behaviour of the two materials.

Orbital projected band structures of FeSe and FeSe/STO [Fig.~\ref{fig:fese_fatbands_comparison}] show that the orbitals associated to the electron pocket at $M$ are Fe $d_{xz}$, $d_{yz}$, and, to a lesser extent, $d_{xy}$. 
In FeSe/STO [Fig.~\ref{fig:fese_sto_fatbands}] a hole pocket associated to interfacial O $p_x$/$p_{y}$ states emerges at $M$.
In the superconducting phase of FeSe/STO, the Fe $d$ states hybridize with O $p_x$ and O $p_y$ and are responsible for the opening of the superconducting gaps at $M$, as predicted by \textit{Acharya} et al.~\cite{acharya2021electronic, acharya2022role, acharya2023vertex}.

\subsection*{Superconducting state}

In order to model the superconducting phase, we solve the
SIESTA--BdG equations using the \textit{fixed}--$\Delta$ approach~\cite{reho2024density}.
We initialize the superconducting pairing potential $\Delta(\mathbf{r})$ in the \textit{superconducting strength representation}. In this approach, the pairing potential is expressed, in real space, via a superconducting strength parameter $\bar{\Delta}$.

\begin{figure}
    \centering
    \includegraphics[width=1.0\columnwidth]{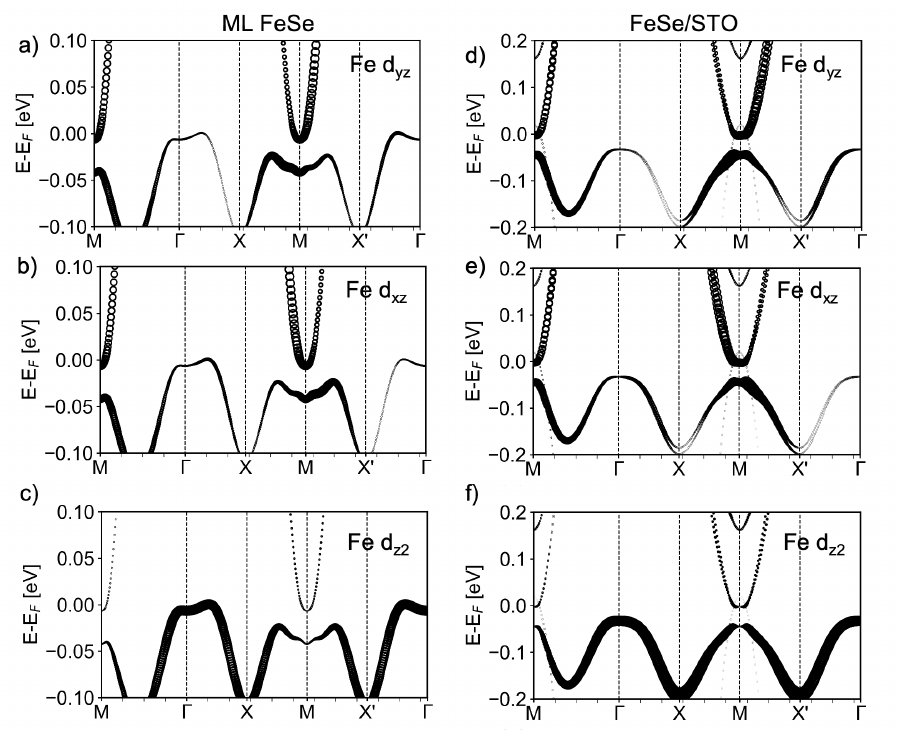}
    \caption{Normal state electronic band structures of (a-c) FeSe ML  and (d-f) FeSe/STO projected on selected Fe orbitals. The projection over all orbitals is reported in Appendix  Figs.~\ref{appfig:fese_ml_fatbands} and ~\ref{fig:fese_sto_fatbands}
}
    \label{fig:fese_fatbands_comparison}
\end{figure}

In ML FeSe, we initialize $\Delta(\mathbf{r})$ with spheres centered on the Fe atoms, each with $\bar{\Delta} = 15$~meV [Fig.~\ref{fig:fese_ml_sc_state}.(a)]. 
The BdG spectrum [Fig.~\ref{fig:fese_ml_sc_state}.(b, c)], the SC--DOS $\rho(\varepsilon)$ and anomalous DOS $\chi({\varepsilon})$ [Fig.~\ref{fig:fese_ml_sc_state}.(d)] reveal several superconducting gaps/coherence peaks 
at energies $\pm 1.8$ ($\Delta_1$), $\pm 5$ ($\Delta_2$), and $\pm 6.4$~meV ($\Delta_3$).
The orbital projected SC--DOS [Fig.~\ref{fig:fese_ml_sc_state}.(f)] shows that the superconducting states are mostly dominated by the Fe $3d_{z^2}$ and, to a minor extent, by $3d_{xz}$ and $3d_{yz}$ orbitals.
We identify the nature of the coherence peaks from the superconducting [Fig.~\ref{fig:fese_ml_sc_state}.(d)]
and orbital-projected normal state (Fig.\ref{fig:fese_fatbands_comparison})
band structures:
The $\Delta_{1}$ coherence peaks originate from Fe $d_{z^2}$ along $X-M$ (close to $M$) and to $d_{xz}/d_{yz}$ along $\Gamma-X$
[Figs.~\ref{fig:fese_fatbands_comparison}.(a, b, c)].
$\Delta_2$ originates from  Fe $d_{xz}$ and $d_{yz}$ along $M-\Gamma$ close to $M$ [Fig.~\ref{fig:fese_fatbands_comparison}.(a, b)].
$\Delta_3$ originates from Fe $d_{z^2}$ along $\Gamma-M$ close to $\Gamma$ [Fig.~\ref{fig:fese_fatbands_comparison}.(c)].
The $\Delta_1$ and $\Delta_2$ peaks exhibit strong electron-hole coupling, as evidenced by the band inversion between electrons (red) and holes (blue)  at the corresponding gap [Fig.~\ref{fig:fese_ml_sc_state}.(d)]. 
From the band structure, we observe a strong anisotropy in the momentum-resolved superconducting gap structure, i.e., $\Delta (\mathbf{k}_1) \neq \Delta(\mathbf{k}_2)$.
We note that the type of Fe orbitals associated with superconductivity and the nature of superconductivity in bulk FeSe differs from ML FeSe:  We have recently showed ~\cite{reho2024density} that the superconducting gap of bulk FeSe is V-shaped, 
and is 
due to Fe $d_{xz}$ and $d_{yz}$ only,  in agreement with experiments \cite{kasahara2014field}. FeSe ML, instead, features a U-shaped gap with multiple coherence peaks, mostly due to $d_{z^2}$ and, in minor part, to $d_{xz}$ and $d_{yz}$ orbitals.

\begin{figure*}
    \centering
    \includegraphics[width=2.0\columnwidth]{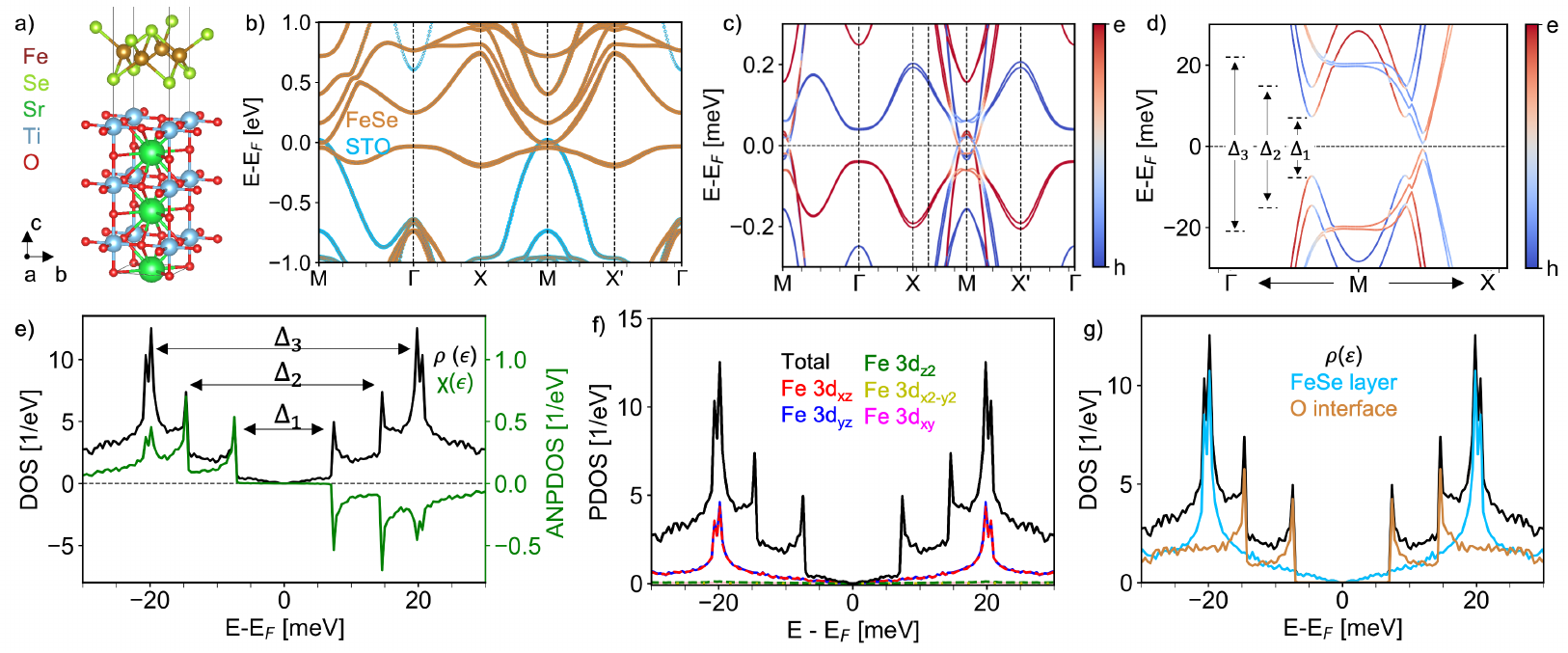}
    \caption{FeSe/STO with a TiO$_2$--terminated surface, FeSe is in the checkerboard AFM magnetic phase: (a) atomistic model (side view). (b) Electronic band structure projected on FeSe (light brown) and STO (light blue).
    (c,d) Superconducting band structure in two different energy ranges. The color scheme highlights contributions from electron-like (red) and hole-like (blue) components of the BdG spectrum. (e) SC--DOS ($\rho(\varepsilon)$, black) and anomalous SC--DOS ($\chi(\varepsilon)$, green) with superconducting gap widths $\Delta_1 = 16.0$, $\Delta_2 = 28.6$, and $\Delta_3 = 40.0$ meV. (f,g) Total (black) and projected SC--DOS over (f) orbitals,  and (g) FeSe layer (light brown) and interfacial O atoms (light blue).
}
    \label{fig:fese_sto_mainplot}
\end{figure*}

FeSe/STO  exhibits three pronounced peaks at $\pm 8$ ($\Delta_1$), $\pm 14.3$ ($\Delta_2$), and $\pm 20$~meV ($\Delta_3$) in the SC--DOS [Fig.~\ref{fig:fese_sto_mainplot}.(e)]. 
The superconducting band structure [Fig.~\ref{fig:fese_sto_mainplot}.(d)] shows several gaps of different sizes near $M$, along the $M-X$ and $M-\Gamma$ directions, demonstrating that $\Delta(\mathbf{k})$ is anisotropic.
The normal-state band structure [Fig.~\ref{fig:fese_sto_mainplot}.(b)] and the projected SC--DOS [Fig.~\ref{fig:fese_sto_mainplot}.(g)] allow us to attribute the $\Delta_1$ and $\Delta_2$ peaks/gaps to the STO substrate and, more specifically, to the interfacial O $p_x$ and $p_y$ states [Figs.~\ref{fig:fese_sto_fatbands}.(g,h)].
Therefore, we deduce that
superconductivity has been induced by proximity on the oxygen atoms.
The $\Delta_3$ peak is, instead, due to Fe and originates along the $M-\Gamma$ line close to $M$.
By comparing with FeSe ML, we deduce that 
the Fe-related SC-gap at $M$ broadens
and changes in orbital character from $d_{z^2}$ to $d_{xz}$ and $d_{yz}$, [Fig.~\ref{fig:fese_sto_mainplot}.f]
in FeSe/STO due to coupling with the substrate. 
The hybridization between FeSe and the substrate, combined with particle-hole symmetry, leads to the emergence of new bands in the BdG spectrum [Figs.~\ref{fig:fese_sto_mainplot}.(c,d)] that are not present in the normal state dispersion [Figs.~\ref{fig:fese_sto_mainplot}.(b)].
These new bands are at $\Gamma$ ($\sim 250$~meV below $E_F$), and $M$ ($\sim 15$--$30$, $70$ and $\sim 180$~meV below $E_F$) originating from the hybridization of Fe  $d_{z^2}$ and O $p_z$ (at 
$\Gamma$) and Fe $d_{xz}$ and $d_{yz}$  with O $p_x$ and $p_y$ (at $M$).
The energy range of these bands corresponds to  the experimentally observed  ``replica bands''~\cite{lee2014interfacial}. 
In our simulations, replica bands originate from: 
(i) the superconducting pairing potential which couples the interfacial region between FeSe and STO, and 
(ii) particle-hole symmetry which mirrors the normal-state conduction states in the valence manifold. 
Our results qualitatively agree with ARPES measurements of FeSe/STO~\cite{liu2015electronic, wang2012interface, huang2017monolayer}.
Differences might be due to the semi-phenomenological nature of the SIESTA--BdG method: 
the initialization of the pairing potential results in an energy shift of the coherence peaks (Appendix~\ref{app:initializationpairingpotential}).

\section*{Discussion}
In this study, we investigated the superconducting properties of ML FeSe,  employing a combined Kohn--Sham and Bogoliubov--de Gennes approach. 
We analyzed two cases: freestanding FeSe (ML FeSe) and FeSe on a TiO$_2$--terminated STO substrate (FeSe/STO). 

For both systems, we predicted an anisotropic superconducting gap $\Delta(\mathbf{k})$ [Figs.~\ref{fig:fese_sto_mainplot}.(d) and~\ref{fig:fese_ml_sc_state}.(d)].
We identified the nature of the superconducting coherence peaks and 
attributed their origin to specific regions in the Brillouin Zone: 
In freestanding FeSe, superconductivity is due to states around $\Gamma$ and $M$ points.
In FeSe/STO, superconductivity originates from states around the $M$ point only. 

In addition, we identified the specific atoms and orbitals that are responsible for superconductivity:
We observed a change in the orbital character of the Fe states responsible for superconductivity from $d_{z^2}$ (ML FeSe) to $d_{xy}/d_{yz}$ (FeSe/STO).
In FeSe/STO, Fe $d$ and O $p$ orbitals strongly hybridize near the Fermi energy at the $M$ point [Figs.~\ref{fig:fese_sto_mainplot}.(b)]
giving rise to two coherence peaks ($\Delta_{1/2}$, mostly contributed by O) and a higher-energy one ($\Delta_{3}$, mostly contributed by Fe).
Remarkably, we observed proximity-induced superconductivity in the interfacial TiO$_2$ layer.

Our analysis of the normal-state band structure of the two materials allowed us to shed light on the different mechanisms behind their superconductivity:
In ML FeSe, metallicity originates from an electron pocket at $M$ and a hole pocket along the $\Gamma-X$ line. In FeSe/STO, instead, the hole pocket shifts to lower energies in the valence band.
Therefore, the superconducting mechanism is intrinsically different in two cases:
In ML FeSe, Cooper pairs can either form at $\sim \Gamma$ and $\sim M$ (difference between electron and hole crystal momentum $Q=0$), or can scatter between $M$ and $\Gamma$ ($Q\neq 0$). 
In FeSe/STO, instead, Cooper pairs formation is only possible at $M$ ($Q=0$).

In both ML FeSe and FeSe/STO, we found multiple competing magnetic configurations. In this work, we focused on the checkerboard antiferromagnetic phase because its band structure best resembles ARPES measurements~\cite{wang2016topological}. 
The impact of different magnetic configurations on the superconducting phase remains an open question, with the possibility that superconductivity may be favored in a specific direction when using the single stripe configuration with different spin alignments along the $x$ and $y$ axes. Future studies could investigate the superconducting state under different magnetic configurations to provide deeper insights into its behavior, or explore the potential for topological superconductivity in FeSe systems coupled with magnetic impurities and strong spin-orbit coupling, such as the recently observed zero energy states at the edges of line defects in Fe(Se, Te)~\cite{chen_atomic_2020}.

Our findings emphasize the role of hybridization with the substrate in determining the superconducting properties of FeSe. 
Hybridization between Fe and the substrate triggers proximity-induced superconductivity in STO.
We suggest that  heterostructures consisting of FeSe and substrates, including elements with high electron affinity, will promote charge transfer and facilitate Cooper pair formation in the substrate, opening the way to engineering novel FeSe-based superconductors operating at high temperatures.

\section*{Methods} 

\subsection*{First-principles ground state simulations}
All simulations were performed using the SIESTA method ~\cite{soler2002siesta,garcia2020siesta},
with the PBE~\cite{perdew1996generalized} functional,  norm-conserving pseudopotentials from the Pseudo-Dojo database~\cite{van2018pseudodojo}, and the default SIESTA double-$\zeta$ polarized (DZP) basis set with an orbital cutoff radii of 272 meV.
Simulations were converged for a mesh--cutoff of 1300~Ry.
In FeSe/STO we account for the correlation effects between Fe and the interfacial oxygen atoms by applying a Hubbard correction U of $0.1$ eV to the Fe $d$ states~\cite{liechtenstein1995density,bousquet2010j}. This value of U allows us to reproduce reasonably the band dispersion observed in ARPES ~\cite{wang2016magnetic,zhang2014interface}, 
in particular the position of the highest occupied state at $\Gamma$, as illustrated in Fig.~\ref{fig:fese_sto_U_tests}).
All calculations include SOC.

To solve the KS-equations, we employed a $101\times101\times1$ and $16\times16\times1$  \textbf{k}-grid for ML FeSe and FeSe/STO, respectively.
The Fermi-Dirac occupation function was smoothed out using an electronic temperature of 10 meV in the normal state for both freestanding FeSe and FeSe/STO. This value has been lowered to $0.1$ meV for the superconducting state simulations to ensure that the broadening of the electronic states is well below the size of the superconducting gap.
The crystal structures were relaxed with a maximum force tolerance of 0.005~eV/\AA.

\subsection*{SIESTA-BdG method}

The superconducting properties were modeled using the \textsc{SIESTA}--BdG method~\cite{reho2024density}. In this approach, a semi-empirical superconducting pairing potential is introduced on top of the self-consistent Kohn-Sham Hamiltonian describing the normal state. 
We used the \textit{fixed--$\Delta$} method with the superconducting pairing potential initialized in real space 
({\it superconducting strength representation})
as touching spherical hardwells. 
We used hardwells with radius $\mathbf{r} =1.32$~\AA~($\mathbf{r} =1.39$~\AA)~and strength $\bar{\Delta} = 15.0$~meV ($\bar{\Delta} = 73.0$~meV) around the Fe atoms for ML FeSe (FeSe/STO).
We used a radius of $\mathbf{r}=1.5$~\AA~and $\bar{\Delta} = 11.0$~meV for the interfacial oxygen atoms.
For a discussion on the initialization of different pairing potential see Supporting Information~\ref{app:initializationpairingpotential}.The superconducting DOS were sampled with a uniform $350\times350\times 1$ and $301\times301\times 1$ \textbf{k}-grid, 
and refined with one iteration of the adaptive-grid scheme with a $3\times3\times1$ subgrid for ML FeSe and FeSe/STO, respectively.

\section*{Acknowledgments}\label{sec:acnkowledgements}
The authors acknowledge the fruitful discussion with Ingmar Swart, Samir Lounis, and Eberhard K. U. Gross.
ZZ acknowledges the research program “Materials for the Quantum Age” (QuMat) for financial support. This program (registration number 024.005.006) is part of the Gravitation program financed by the Dutch Ministry of Education, Culture and Science (OCW).
RR and AK acknowledges financial support from Sector Plan Program 2019-2023. 
NW acknowledges support from the EU MaX CoE (Grant No. 101093374) and Grants No. PCI2022-134972-2 and PID2022-139776NB-C62 funded by the Spanish MCIN/AEI/10.13039/501100011033 and by the ERDF, A way of making Europe.  
NW further acknowledges funding from  the  European  Union’s  Horizon  2020  research  and  innovation  programme  under  the  Marie  Skłodowska-Curie  Grant  Agreement  No.  754558  (PREBIST  –  COFUND).  
ICN2 is supported by the Severo Ochoa programme from Spanish MINECO (Grant no. CEX2021-001214-S) and by Generalitat de Catalunya (CERCA program and Grant No. 2021SGR01519).
This work was sponsored by NWO-Domain Science for the use of supercomputer facilities. We also acknowledge that the results of this research have been achieved using the Tier-0 PRACE
Research Infrastructure resource Discoverer based in Sofia, Bulgaria  (OptoSpin project id. 2020225411). 
This project has received funding from the European Union’s Horizon Europe research and innovation program under Grant Agreement No 101130384 (QUONDENSATE).

\section*{Data availability statement}
The data that support the findings are available upon request. 

\section*{Conflict of Interest}
The authors declare no conflict of interest.

\section*{Code availability}
The \textsc{SIESTA} code is published as open source under the the GNU General Public License on the official \href{https://gitlab.com/siesta-project/siesta}{GitLab repository} while the \textsc{SIESTA}-BdG code is available upon request.

\section*{Keywords}
superconductivity, semiconductor-superconductor heterostructure, BdG equations, quantum materials

\appendix
\section{Magnetic order}\label{app:magnorder}
\begin{figure}[ht]
    \centering
    \includegraphics[width=\columnwidth]{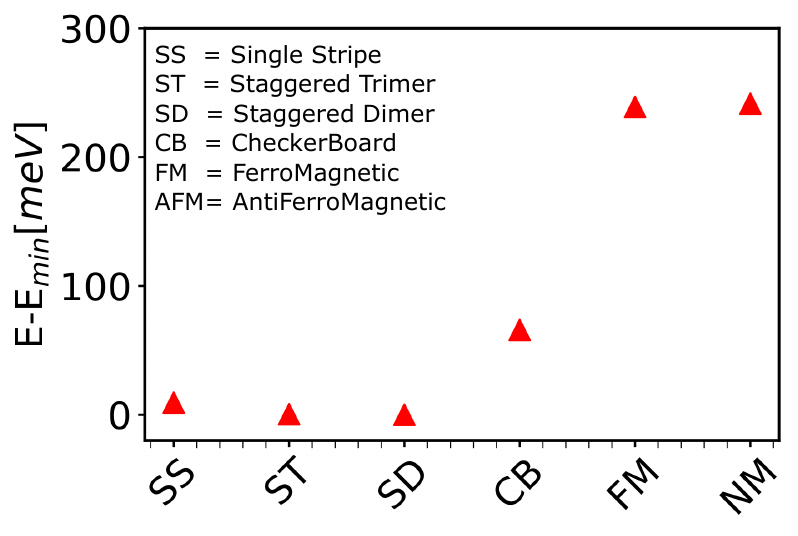}
    \caption{
    Energetic magnetic ordering for FeSe bulk. The lowest energy magnetic state is the Staggered Dimer configuration.
    }
    \label{fig:bulkmagneticorder}
\end{figure}

\begin{figure}[ht]
    \centering
    \includegraphics[width=\columnwidth]{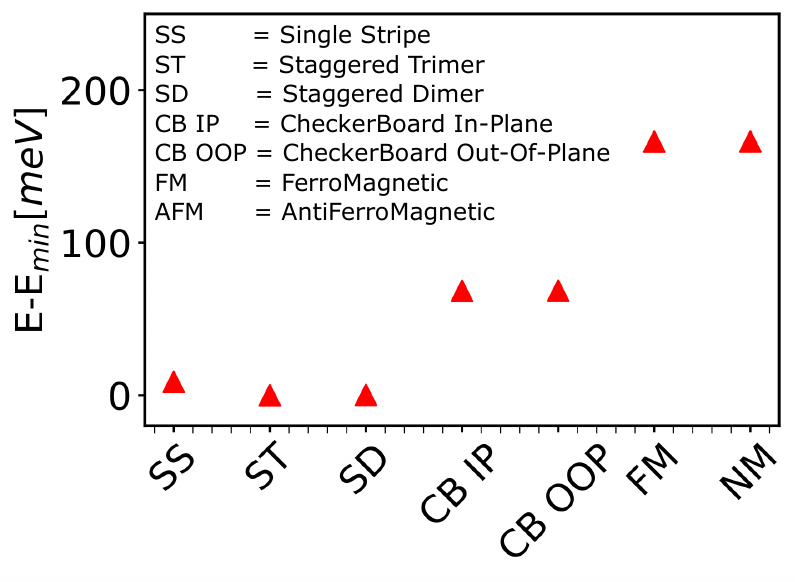}
    \caption{
    Energetic magnetic ordering for ML FeSe. 
    }
    \label{fig:fesemlmagnorder}
\end{figure}

Over the last decade, the ground state of FeSe has been hotly debated in the literature. Magnetic frustration and spin fluctuations make it difficult to describe this system with the Heisenberg or Ising models.
Recently, \textit{Glasbrenner et al.}~\cite{GlasbrennerEffectMagnetic2015} performed a detailed analysis, based both on first principles and model calculations establishing  the Staggered Dimer (SD) as the most stable configuration for this system.
Our first-principles \textit{ab-initio} analysis confirms their result. In Fig.~\ref{fig:bulkmagneticorder} we show the computed total energies for the Single Stripe (SS), Staggered Trimer (ST), Staggered Dimer,  in-plane (IP) and out-of-plane (OOP) CheckerBoard (CB), FerroMagnetic (FM) and non magnetic (NM) configurations. 
We report the difference in total energies $\Delta E$ with respect to the ground state ($E_{min}$) in Fig.s~\ref{fig:bulkmagneticorder} and~\ref{fig:fesemlmagnorder} and Table~\ref{tab:magnenergies}.

Two different basis sets were tested: Double-Zeta Polarized (DZP) with orbital-confining cutoff radii (PAO.EnergyShift) of 272~meV and 25~meV. The first choice leads to physically correct energetic ordering, and therefore our calculations are performed with this basis set.

The low-energy magnetic configurations for ML FeSe are ST, SS, and SD, with the ST only 0.231~meV below the SD. This energy difference is below the 0.5~meV/Fe accuracy of our simulations, hence, we conclude that the three configurations are iso--energetic.  
We report our results in Fig.~\ref{fig:fesemlmagnorder} and Table~\ref{tab:magnenergies}.

\begin{table}[h]
    \caption{\label{tab:magnenergies} 
    Difference in total energies $\Delta E = E - E_{\m{min}}$, of ML FeSe and FeSe/STO with different magnetic configurations. The minimal energies are highlighted in bold.
    }
\centering
\begin{tabular}{l|c}
\hline
\textbf{ML FeSe} & \textbf{$\Delta E [meV]$} \\
\hline
SS & 8.868 \\
\textbf{ST} & \textbf{0.000} \\
SD & 0.231 \\
CB IP & 68.453 \\
CB OOP & 68.557 \\
FM & 166.314 \\
NM & 166.357 \\
\hline
\textbf{Bulk FeSe} & \textbf{$\Delta E [meV]$} \\
\hline
SS & 9.418 \\
ST & 0.424 \\
\textbf{SD} & \textbf{0.000} \\
CB & 65.846 \\
FM & 238.971 \\
NM & 241.511 \\
\hline
\end{tabular}
\end{table}

\section{Initialization of the Pairing Potential and Its Impact on the Superconducting Properties}\label{app:initializationpairingpotential}
In the \textsc{SIESTA}-BdG method, one can employ different approaches for solving 
the BdG equations and initializing the pairing potential $\Delta$~\cite{reho2024density}. 
We explored multiple initialization schemes for the pairing potential in both ML FeSe and FeSe/STO.
In the main text, we focus on results obtained using the
\textit{fixed--$\Delta$}  method to solve the \textsc{SIESTA}-BdG equations, initialized within the 
\textit{superconducting strength representation}.
The \textit{fixed--$\Delta$} algorithm solves the Bogoliubov--de Gennes equations ensuring self-consistency in the normal Hamiltonian and density while holding $\Delta$ constant during the \textsc{SIESTA}--BdG scf steps. At every scf step the normal state Hamiltonian, the normal and the anomalous densities are updated. 
The \textit{superconducting strength representation} initializes $\Delta$ in real space by specifying an initial guess for the pairing strength, $\bar{\Delta}$ (when no risk of confusion arise, we simply write $\Delta$).

\begin{figure}[hb]
    \centering
    \includegraphics[width=1.\linewidth]{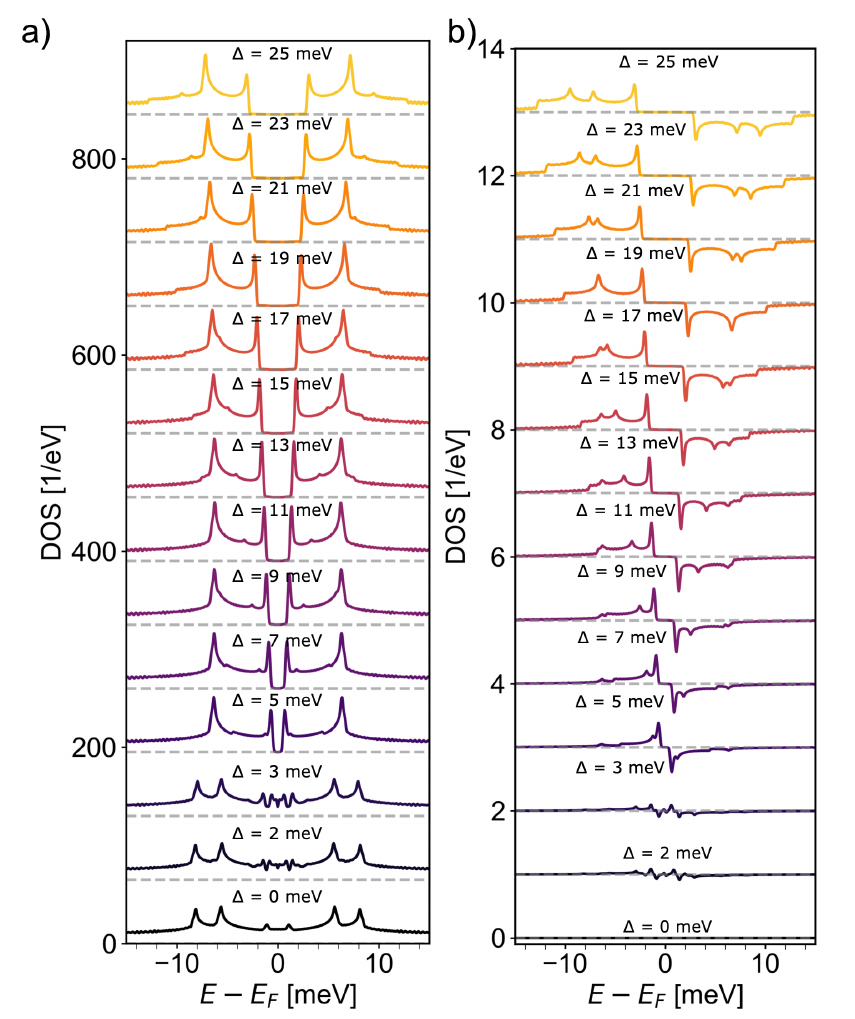}
    \caption{(a) SC--DOS and (b) anomalous DOS (right) for ML FeSe varying the initial value of $\Delta$ from 0 to 25~meV. The superconducting pairing potential is initialized only on the Fe atoms.}
    \label{fig:fese_ml_varying_delta}
\end{figure}

\begin{figure}[h]
    \centering    
    \includegraphics[width=1.\linewidth]{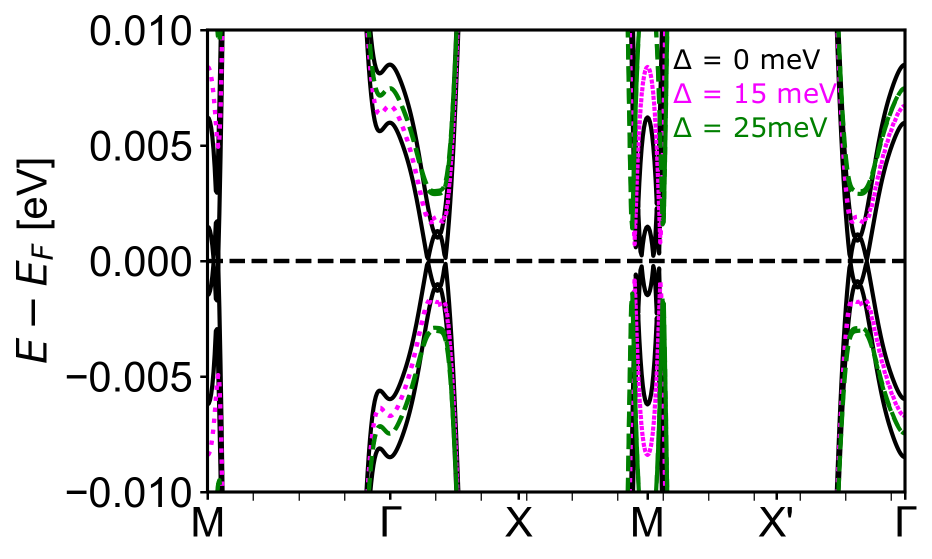}
    \caption{Superconducting band structure of ML FeSe with $\Delta$ = 0 (black), 15 (magenta), or 25 (green) meV. The superconducting pairing potential is initialized only on the Fe atoms.}
    \label{fig:fese_ml_bands_varying_delta}
\end{figure}

For ML FeSe, we analysed the shape and magnitude of the SC--DOS $\rho(\varepsilon)$ while varying the initial value of $\bar{\Delta}$.
As shown in Fig.~\ref{fig:fese_ml_varying_delta}, $\rho(\varepsilon)$ is zero for $\bar{\Delta}=0$, indicating the absence of superconductivity. 
A transition from a non-gapped to fully gapped SC--DOS occurs as $\bar{\Delta}$ increases from $3$ to $5$~meV. 
Namely,  $\bar{\Delta} < 5$ meV is too small to describe superconductivity in ML FeSe.
For $\bar{\Delta}=5$~meV, the double peaks around $\pm 8$~meV merge into a single peak. Further increases in $\bar{\Delta}$ result in broader coherence peaks, reflecting the enhanced value of the superconducting pairing potential. 
Experimentally, different values for the superconducting gap width  have been reported, with $\bar{\Delta} = 15$ meV in~\cite{liu2015electronic}. 
The BdG spectrum and SC--DOS (Figs.~\ref{fig:fese_ml_bands_varying_delta} and~\ref{fig:fese_ml_varying_delta}) does not, qualitatively, change for $\bar{\Delta}$ close to $15$ meV. Therefore, we employed $\bar{\Delta} = 15$ meV for the \textsc{SIESTA}-BdG simulations presented in the main text.

\begin{figure}[h]
    \centering
    \includegraphics[width=1.\linewidth]{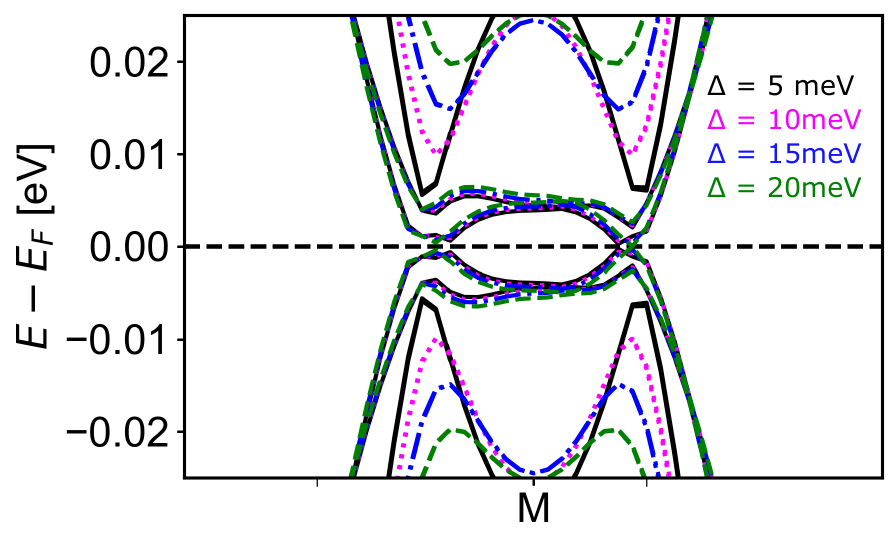}
    \caption{Superconducting band structure of FeSe/STO with $\Delta$ = 0 (black), 10 (magenta), 15 (blue), or 20 (green) meV. The superconducting pairing potential is initialized on the Fe atoms and the oxygen atoms of the TiO$_2$ terminated surface.}
    \label{fig:fese_sto_scbands_varying_delta}
\end{figure}

\begin{figure}[h]
    \centering
    \includegraphics[width=1.\linewidth]{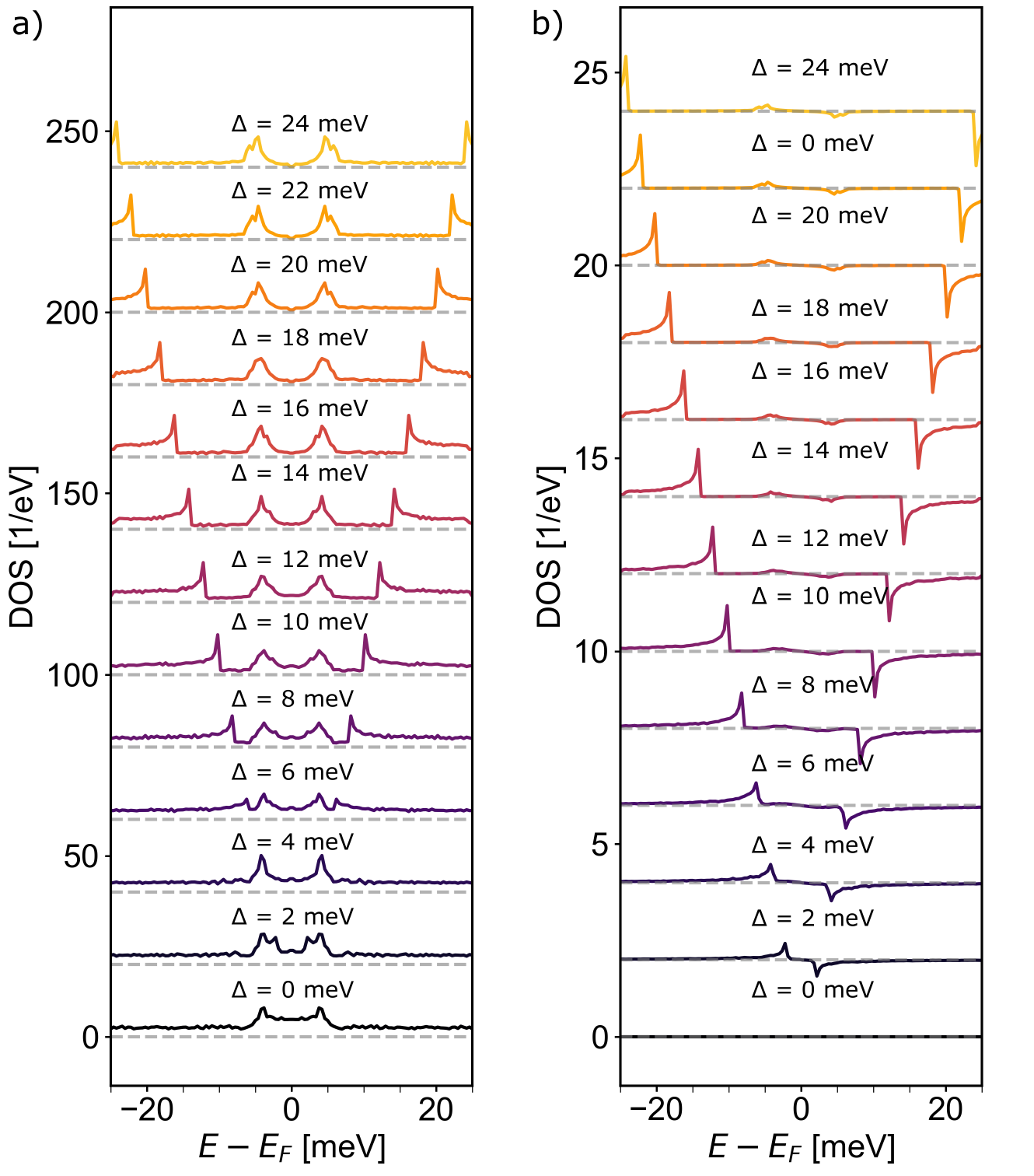}
    \caption{(a) SC--DOS and (b) anomalous DOS (right) for FeSe/STO obtained by varying $\Delta$ from 0~meV to 24~meV. The superconducting pairing potential is initialized in the whole unit cell with a constant value.}
    \label{fig:fese_sto_varying_constant_delta}
\end{figure}

\begin{figure}[h]
    \centering
    \includegraphics[width=1.\linewidth]{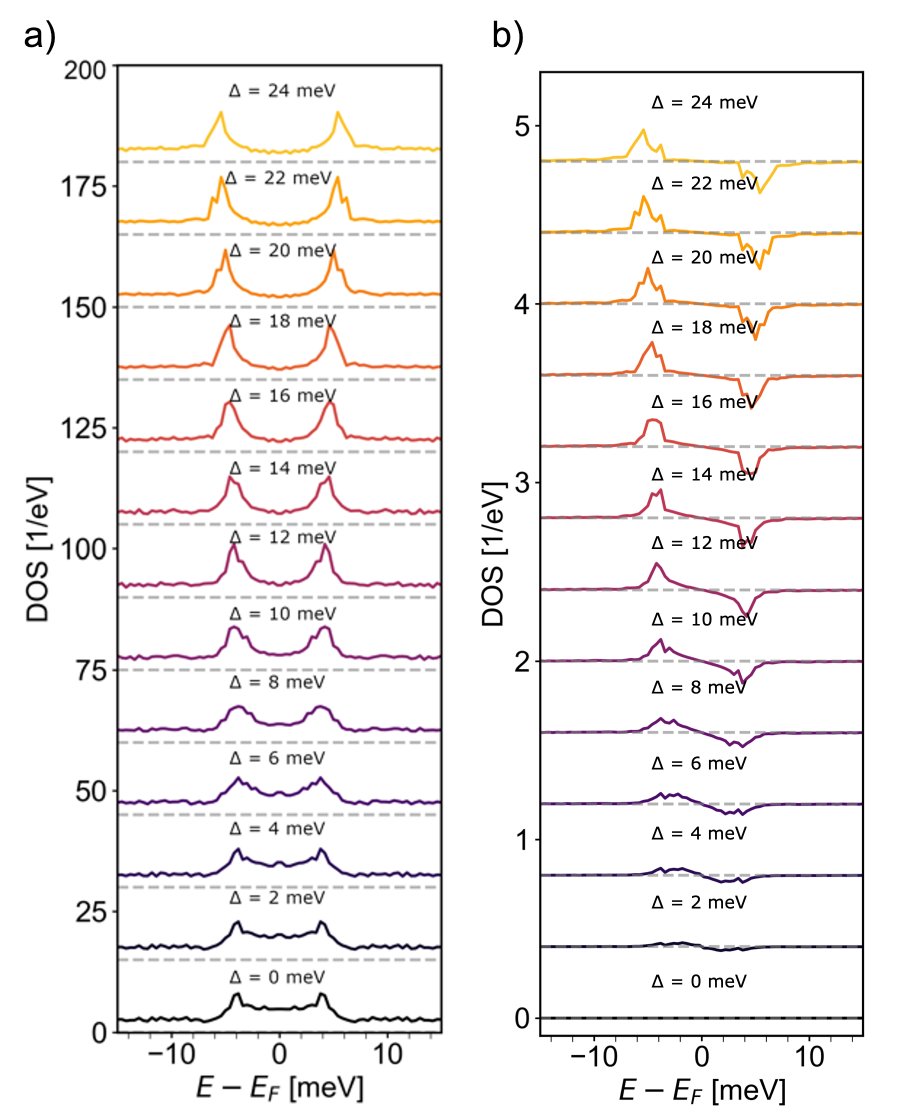}
    \caption{(a) SC--DOS and (b) anomalous DOS (right) for FeSe/STO by obtained by varying $\Delta$ from 0 to 24~meV. The superconducting pairing potential is initialized only on the Fe atoms.}
    \label{fig:fese_sto_varying_fe_delta}
\end{figure}

For FeSe/STO, the interpretation of the SC--DOS is more complex. Our results indicate that only the states near the FeSe/STO interface contribute significantly to $\rho(\varepsilon)$ close to the Fermi level, regardless of the spatial initialization of the pairing potential.
In order to assess this stament, we first initialize the pairing potential $\bar{\Delta}$ uniformly across the unit cell (Fig.~\ref{fig:fese_sto_varying_constant_delta}) with values ranging from 0 to 24~meV. The SC--DOS is metallic for $\bar{\Delta}=0$. As $\bar{\Delta}$ increases , two peaks are clearly distinguishable and their separation  widens.
At $\bar{\Delta} = 0$~meV, we observe a pair of coherence peaks at $\pm 4.5$~meV, which we attribute to the gaps in flat degenerate bands near the Fermi level in the normal state of FeSe/STO [Fig.~\ref{fig:fese_sto_mainplot}.(b)]. These peaks primarily orginate from Fe $d$ orbitals. As $\bar{\Delta}$ increases, a second pair of coherence peaks emerges starting at $\bar{\Delta} = 6$~meV. This second pair, associated with interfacial O $p$--orbitals, becomes broader with increasing $\bar{\Delta}$, while the first pair remains unchanged.
Thus, the constant initialization of the pairing potential indicates that both Fe and oxygen play a role in the superconducting coupling, although the exact peak positions are not quantitatively accurate.
Furthermore, the SC--DOS does not completely drop to zero regardless of the value of $\bar{\Delta}$ while the anomalous DOS does. 
We interpret this behavior as follows: the first pair of coherence peaks arises from intrinsic electronic properties (coupling to STO), while the second is driven by the superconducting pairing potential, as evidenced by the trends in $\chi(\varepsilon)$.
The BdG spectrum near the high--symmetry point $M$ (Fig.~\ref{fig:fese_sto_scbands_varying_delta}) provides further insights. Along the $M-X^\prime$ path (left) and the $M-X$ path (right), we observe a widening separation between electron and hole states as $\bar{\Delta}$ increases, except for states near the Fermi level. Notably, a gap--closing feature appears along the $M-X^\prime$ path, which is absent along $M-X$. This suggests a breaking of inversion symmetry in the system.
A value of $\bar{\Delta}=24$~meV is not sufficient to fully open the gap for FeSe/STO. Our tests indicate that at least a value of $45$~meV is needed. 
For the reasonings above, we initialized the pairing potential $\bar{\Delta}$ only on the Fe atoms of the ML and the interfacial oxygen atoms.
We tested that, qualitatively, no differences arise if the superconducting pairing potential is initialized uniformly across the unit cell or only on the Fe and interfacial Oxygen atoms.
We also tested the case where the superconducting pairing potential is initialized as touching spheres on the Fe atoms (Fig.~\ref{fig:fese_sto_varying_fe_delta}).

\begin{figure}[h]
    \centering
    \includegraphics[width=1.\linewidth]{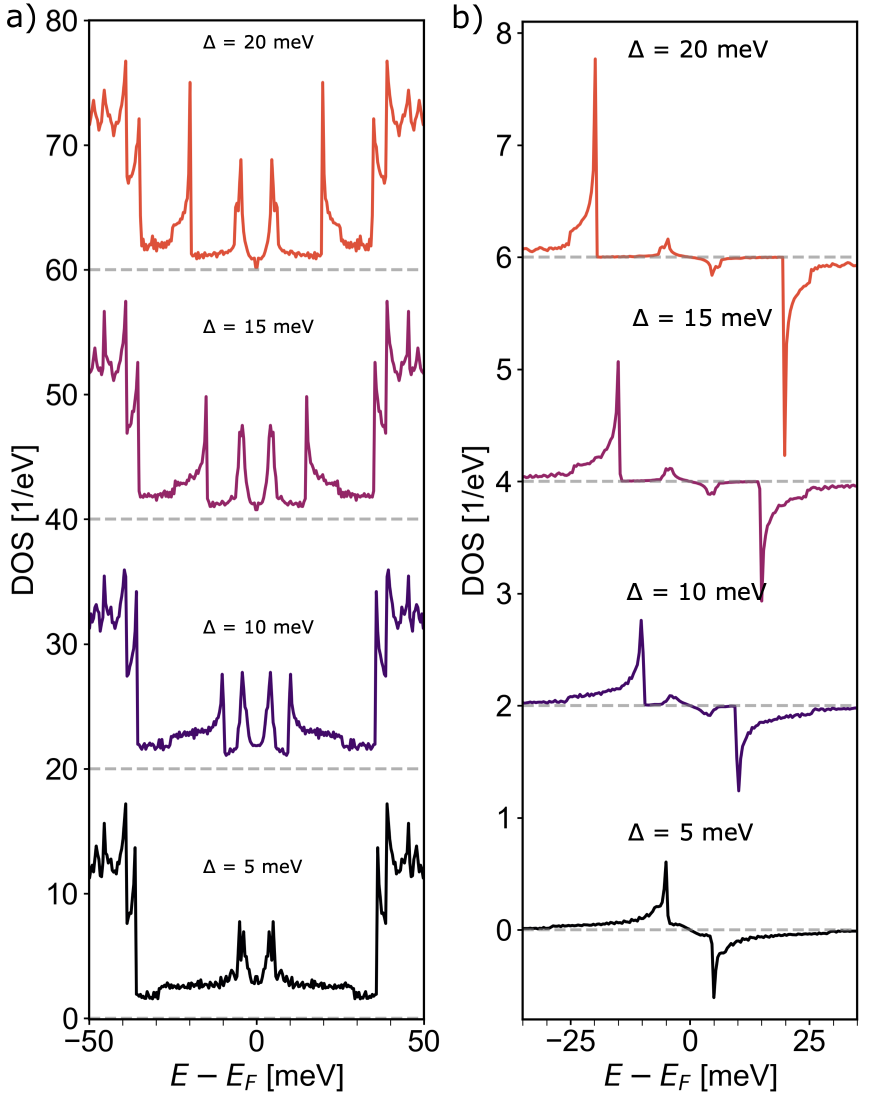}
    \caption{(a) SC--DOS and (b) anomalous DOS (right) for FeSe/STO varying $\Delta$ from 0~meV to 24~meV. The superconducting pairing potential is initialized only on the Fe atoms and the oxygen atoms of the TiO$_2$ terminated surface.}
    \label{fig:fese_sto_varying_feo_delta}
\end{figure}

\begin{figure}[h]
    \centering
    \includegraphics[width=1.\linewidth]{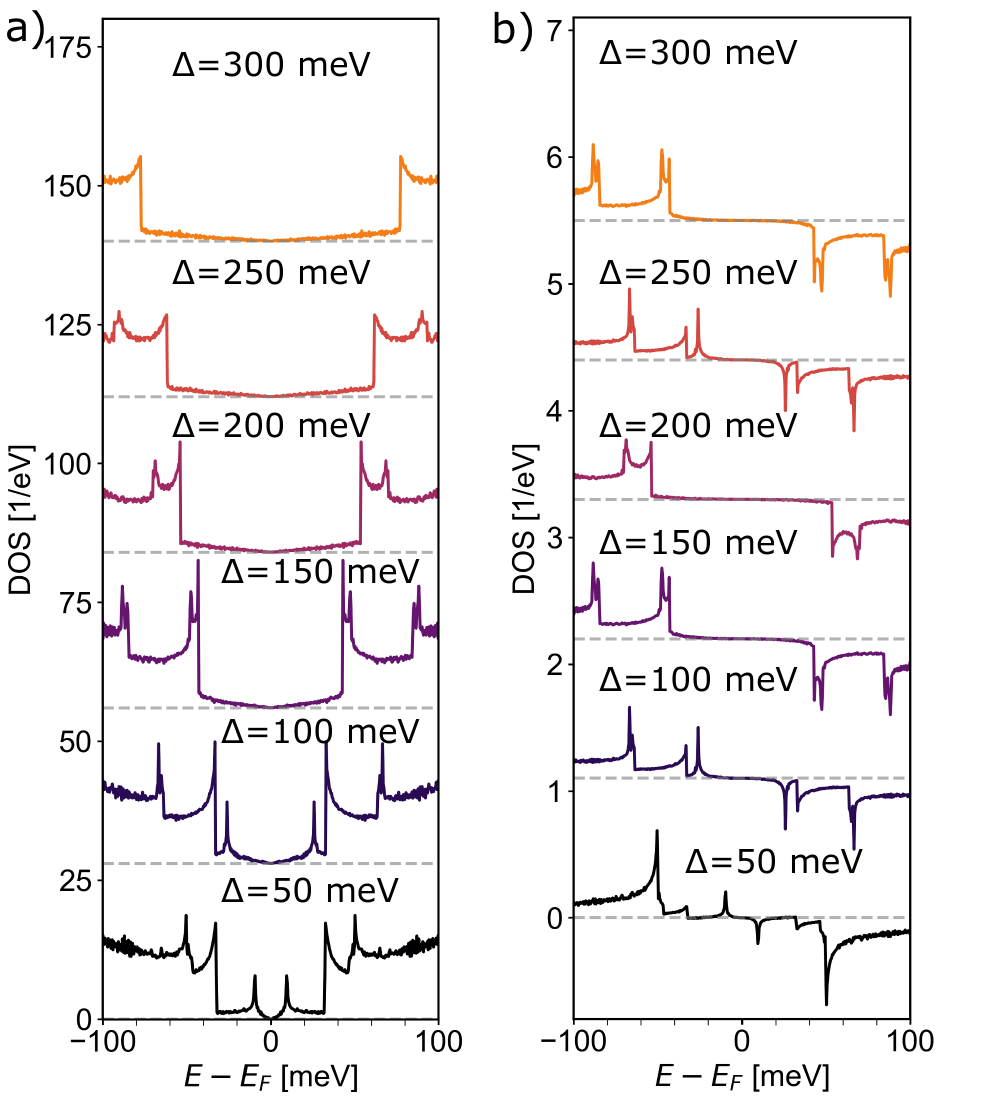}
    \caption{(a) SC--DOS and (b) anomalous DOS (right) for FeSe/STO varying $\Delta$ from 0~meV to 24~meV. The superconducting pairing potential is initialized only on the Fe atoms and the oxygen atoms of the TiO$_2$ terminated surface.}
    \label{fig:fese_sto_varying_feo_delta2}
\end{figure}

For FeSe/STO, we report the SC--DOS and anomalous DOS for different values of $\bar{\Delta}$ from 0~meV up to $300$~meV, with the superconducting pairing initialized on the Fe interfacial oxygen atoms (Fig.s~\ref{fig:fese_sto_varying_feo_delta} and~\ref{fig:fese_sto_varying_feo_delta2}).
These results show that for a higher value $\bar{\Delta}$ the SC--DOS becomes gapped.

\section{Supplementary Electronic Structure Analysis of ML FeSe and FeSe/STO}
\begin{figure}[h]
    \centering
    \includegraphics[width=1.\linewidth]{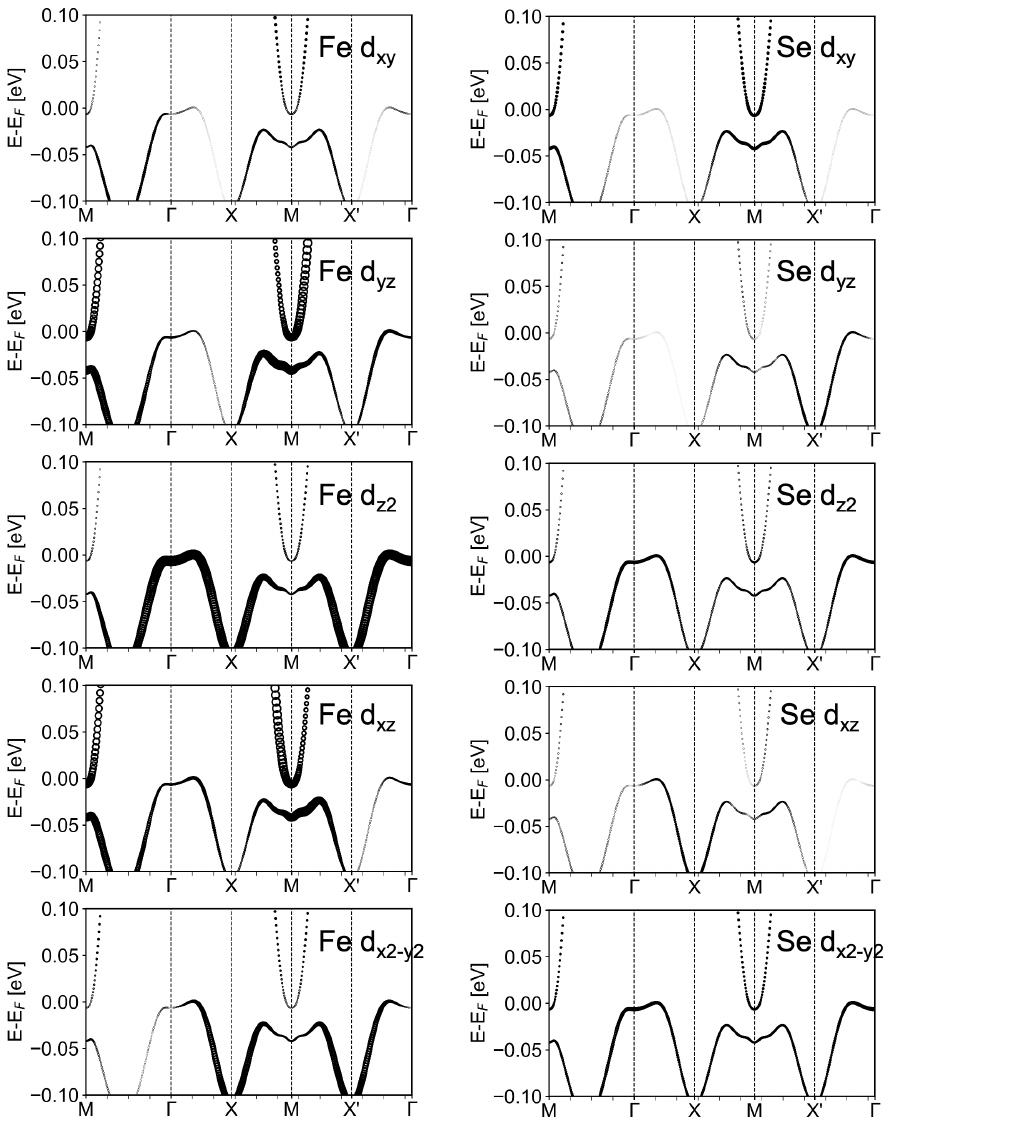}
    \caption{Electronic band structure of ML FeSe, projected on Fe(left) or Se(right) $d$ orbitals.}
    \label{appfig:fese_ml_fatbands}
\end{figure}

\begin{figure}[h]
    \centering
    \includegraphics[width=1.\linewidth]{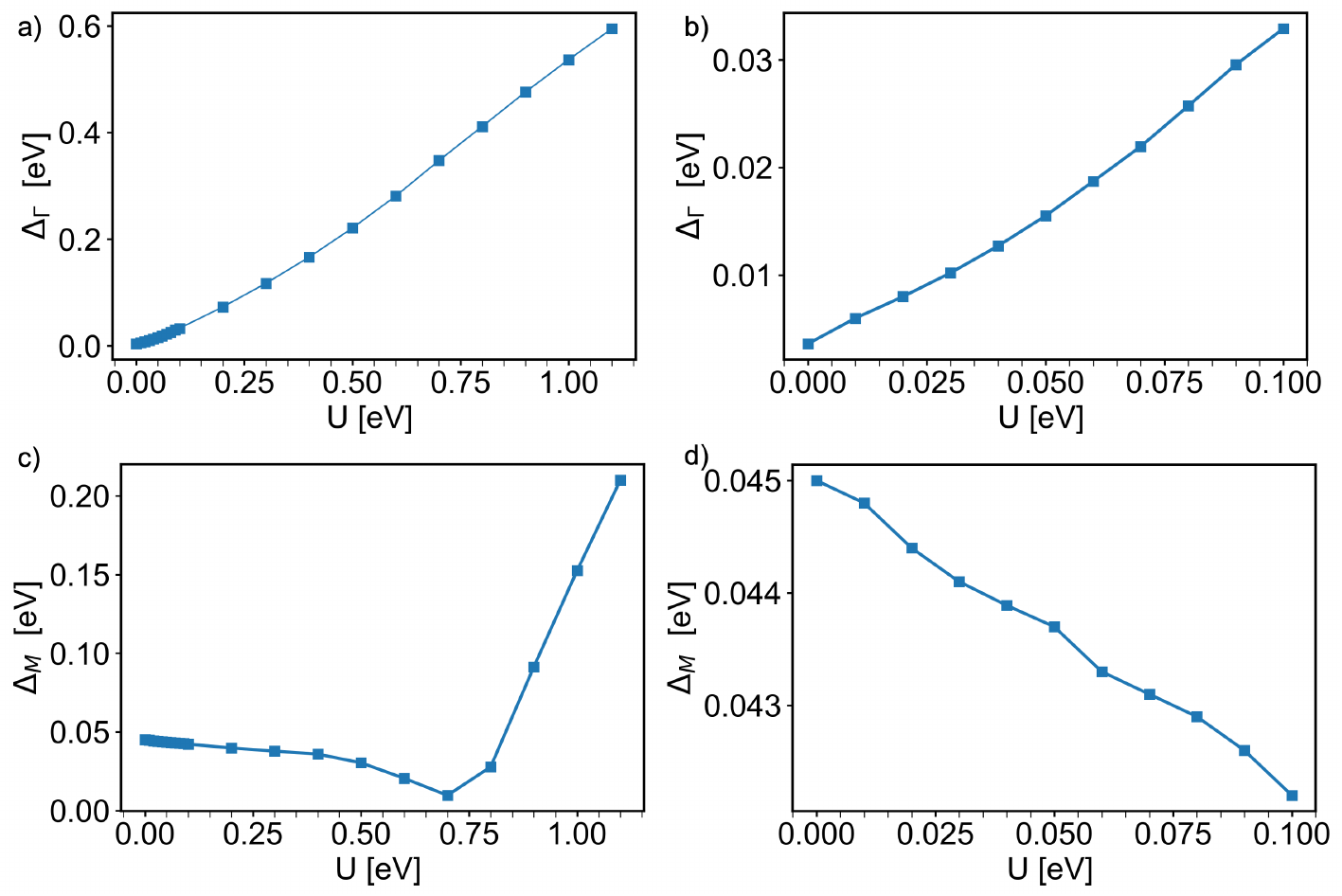}
    \caption{Convergence of the Hubbard U parameter towards ARPES results: (a) Energy difference between
     $E_F$ and the last occupied state at $\Gamma$, $\Delta_{\Gamma}$ (b zoom-in) and (c,d) $M$, $\Delta_M$, as a function of the Hubbard U term.}
    \label{fig:fese_sto_U_tests}
\end{figure}

\begin{figure*}[t]
    \centering
    \includegraphics[width=1.\linewidth]{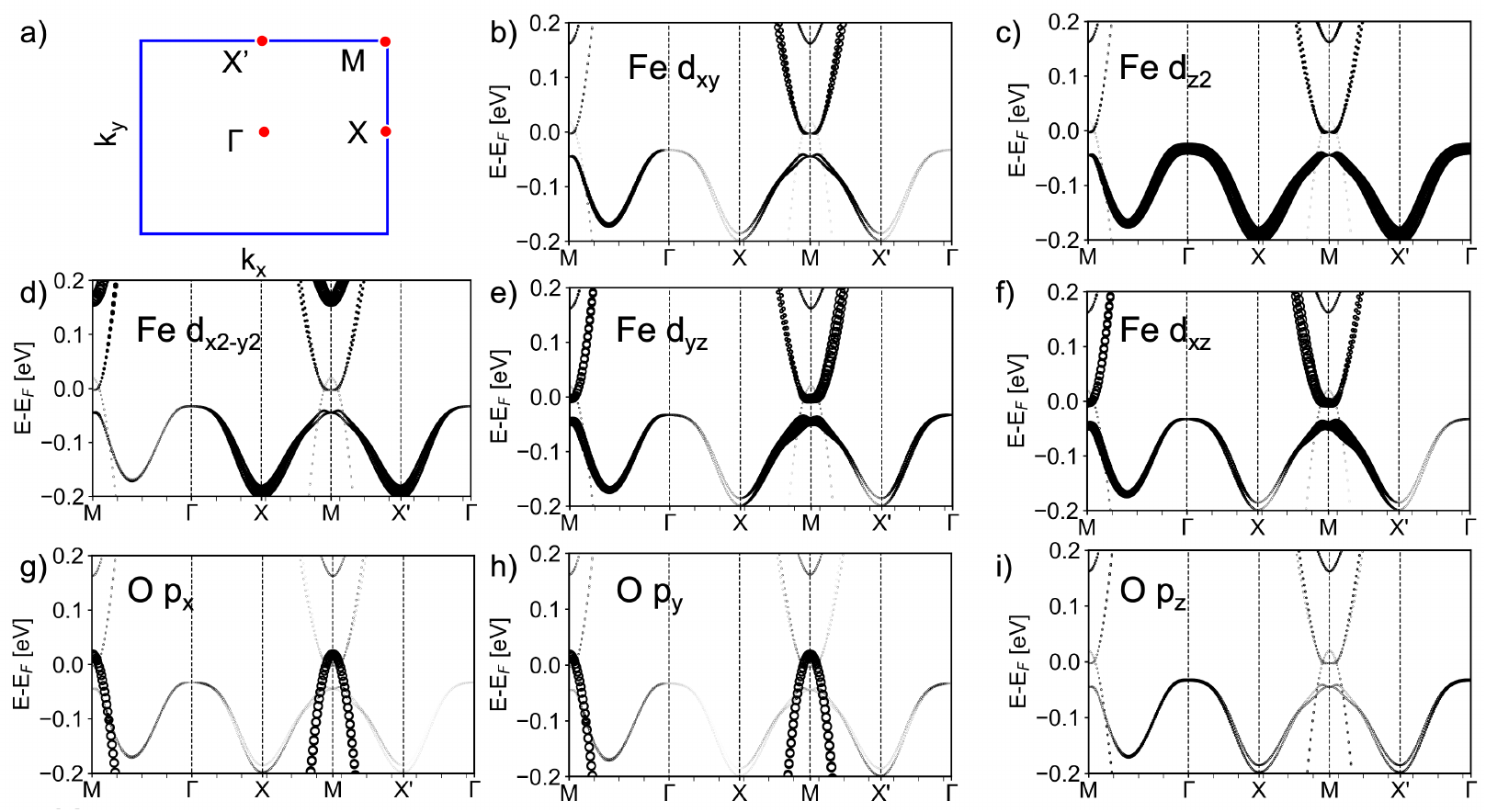}
    \caption{FeSe/STO:
     Schematic representation of the Brillouin Zone and its high-symmetry points. (a)
    Normal state electronic band structure projected on  (b--f) $d$ orbitals of Fe  and  (g--i) $p$--orbitals of O .}
    \label{fig:fese_sto_fatbands}
\end{figure*}

\begin{figure}[h]
    \centering
    \includegraphics[width=1.\columnwidth]{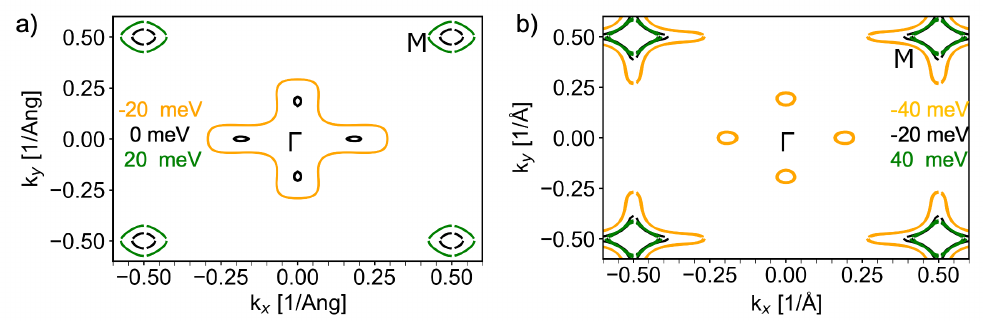}
    \caption{Fermi surfaces of (a) FeSe ML and (b) FeSe/STO in the normal state.
    The surfaces are shown as constant-energy cuts in the (E, $k_x$, $k_y$) frame. Energy values and corresponding colors are indicated in the legend.}
    \label{fig:fermi_surfaces}
\end{figure}

In this section, we present supplementary figures that support the discussion in the main text. We compute the orbital-projected band structure of ML FeSe (Fig.~\ref{appfig:fese_ml_fatbands}) and FeSe/STO (Fig.~\ref{fig:fese_sto_fatbands}). 
We show the Fermi surfaces of ML FeSe and FeSe/STO (Fig.~\ref{fig:fermi_surfaces}).
We examine the changes in the electronic dispersion of FeSe/STO as a function of the Hubbard U parameter (Fig.~\ref{fig:fese_sto_U_tests}). We compare our results with ARPES data~\cite{liu2015electronic} and employ a value of $U=0.1$~eV for the simulations performed in the main text. 


\clearpage
\bibliographystyle{unsrt}
\bibliography{bibliography}

\end{document}